\documentclass[prl,aps,amsmath,amssymb,superscriptaddress,twocolumn,hypertext]{revtex4-1}

\usepackage{graphicx}
\usepackage{dcolumn}
\usepackage{bm}
\usepackage{xcolor}
\usepackage[latin9]{inputenc}
\usepackage{amsmath}
\usepackage{multirow}
\usepackage[per-mode=symbol]{siunitx}

\DeclareSIUnit\gauss{G}
\graphicspath{{/}}

{\tiny }

\begin{document}


\title{Imaging and localizing individual atoms interfaced with a nanophotonic waveguide}

\author{Y.~Meng}
\affiliation{%
	Vienna Center for Quantum Science and Technology,\\
	TU Wien -- Atominstitut, Stadionallee 2, 1020 Vienna, Austria
}%
\affiliation{%
 Department of Physics, Humboldt-Universit\"at zu Berlin, 10099 Berlin, Germany
}%
\author{C.~Liedl}
\affiliation{%
	Department of Physics, Humboldt-Universit\"at zu Berlin, 10099 Berlin, Germany
}%
\author{S.~Pucher}

\affiliation{%
	Vienna Center for Quantum Science and Technology,\\
	TU Wien -- Atominstitut, Stadionallee 2, 1020 Vienna, Austria
}%
\affiliation{%
	Department of Physics, Humboldt-Universit\"at zu Berlin, 10099 Berlin, Germany
}%
\author{A.~Rauschenbeutel}
\affiliation{%
	Vienna Center for Quantum Science and Technology,\\
	TU Wien -- Atominstitut, Stadionallee 2, 1020 Vienna, Austria
}%
\affiliation{%
 Department of Physics, Humboldt-Universit\"at zu Berlin, 10099 Berlin, Germany
}%
\author{P.~Schneeweiss}
\email{philipp.schneeweiss@hu-berlin.de}
\affiliation{%
	Vienna Center for Quantum Science and Technology,\\
	TU Wien -- Atominstitut, Stadionallee 2, 1020 Vienna, Austria
}%
\affiliation{%
	Department of Physics, Humboldt-Universit\"at zu Berlin, 10099 Berlin, Germany
}%

\date{\today}

\begin{abstract}
Single particle-resolved fluorescence imaging is an enabling technology in cold-atom physics. However, so far, this technique was not available for nanophotonic atom--light interfaces. Here, we image single atoms that are trapped and optically interfaced using an optical nanofiber. Near-resonant light is scattered off the atoms and imaged while counteracting heating mechanisms via degenerate Raman cooling. We detect trapped atoms within \SI{150}{\milli\second} and record image sequences of given atoms. Building on our technique, we perform two experiments which are conditioned on the number and position of the nanofiber-trapped atoms. We measure the transmission of nanofiber-guided resonant light and verify its exponential scaling in the few-atom limit, in accordance with Beer-Lambert's law. Moreover, depending on the interatomic distance, we observe interference of the fields that two simultaneously trapped atoms emit into the nanofiber. The demonstrated technique enables post-selection and possible feedback schemes and thereby opens the road towards a new generation of experiments in quantum nanophotonics. 
\end{abstract}

\maketitle

Engineering light-matter interaction at the level of single atoms and photons is one of the main pursuits in quantum optics. Over the last decade, various techniques have been developed for trapping and optically interfacing atoms using nanophotonic devices~\cite{Vetsch10,Thompson13b,Goban15,Kato15,Lee15,Nayak19}. By tailoring the nanophotonic mode structure, e.g., via bandgaps or resonances, the interaction between the atoms and the mode can be engineered and enhanced. Therefore, nanophotonic systems offer intriguing and often unique opportunities~\cite{Chang18}. For example, they can be used to study strong light-matter coupling in an integrated setting~\cite{Thompson13b,Kato15} and in unconventional parameter regimes~\cite{Dareau18,Johnson19}. Moreover, chiral light--matter interaction occurs naturally near nanophotonic structures~\cite{Lodahl17}, and several building blocks for future quantum networks have already been demonstrated using nanophotonic, fiber-integrated cold-atom systems~\cite{Nayak09,Sayrin15a,Gouraud15,Corzo19}.

Recently, the dynamics of the number of nanofiber-trapped atoms in a large ensemble has been measured using heterodyne detection~\cite{Beguin14}, and the preparation of atoms in the motional ground state of a nanofiber-based trap has been demonstrated~\cite{Meng18}. Two important next steps in order to further enhance the control of atoms near nanophotonic structures are to image and to address the atoms individually. For free-space optical tweezers and lattices, many remarkable scientific results have been enabled by such techniques~\cite{Bergamini04,Meschede06, Bakr09,Weitenberg11,Cheuk15,Bernien17,Barredo18}. Imaging and addressing single atoms is even more challenging in nanophotonic cold-atom systems: There, scattering of the excitation light by the nearby nanophotonic structure hampers the detection of the fluorescing atoms. Moreover, Raman-scattering of trapping laser fields in the waveguide material can produce additional near-resonant background light. Recently, imaging of a single trapped atom that is placed in close vicinity of but not yet coupled to a nanophotonic circuit has been demonstrated~\cite{Kim19}. In addition, images of atoms that are trapped in optical tweezer arrays at a distance of \SI{\sim 3}{\milli\meter} from a photonic crystal waveguide have been recorded~\cite{Beguin20}. Furthermore, the fluorescence of a single trapped atom has been detected with a nanofiber-based cavity~\cite{Nayak19}.

In this work, we demonstrate imaging of individual atoms that are trapped and optically interfaced using the evanescent field surrounding an optical nanofiber~\cite{Vetsch10}. The imaging is performed by means of light that the atoms emit during degenerate Raman cooling (DRC)~\cite{Lester14}. The cooling prepares the atoms both in a well-defined internal state and keeps them close to the motional ground state of the trap. Our imaging capability then allows us to post-select experimental runs on the number of trapped atoms. In this way, we determine the transmission of a resonant light field through the nanofiber as a function of the atom number. This allows us to directly measure the extinction per nanofiber-coupled atom. Moreover, by measuring the power of the light that two trapped atoms emit into the nanofiber, we observe interference as a function of the atom--atom distance.

\begin{figure}
	\includegraphics[width=1\columnwidth]{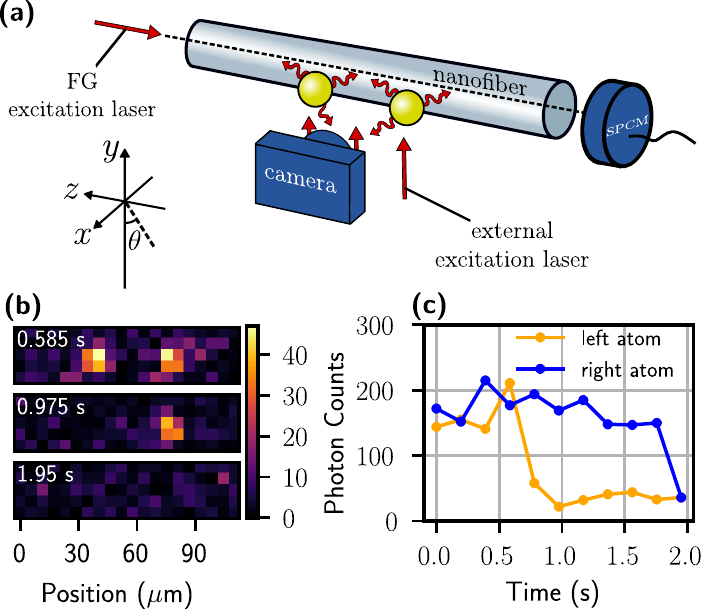}
	\caption{\textbf{Experimental scheme and atom imaging}. (a) A few cesium atoms are trapped $\sim \SI{270}{\nano\meter}$ away from the surface of an optical nanofiber. They are exposed to a near-resonant excitation laser field that freely propagates along the $+y$-direction. The atoms scatter this light and are imaged using a camera. A fraction of the scattered photons also couples to the nanofiber-guided mode. The atoms can also be excited or probed with light launched into the nanofiber. The light exiting the nanofiber can be detected with a single photon counting module (SPCM). (b) Typical raw images of two, one, and zero atoms for an exposure time of \SI{150}{\milli\second}. (c) Counts summed over regions of $3\times 3$ pixels, centered on the atoms shown in (b) as a function of time. A 45-ms waiting time between consecutive images is reserved. The count rates drop to the background level when the atoms are lost from the trap after 0.975~s (left atom) and 1.95~s (right atom). See supplementary material (SM) for the full series of images underlying the plot.
}
	\label{fig:ExpSetup_combined}
\end{figure}

The core elements of the experiment are shown in Fig.~\ref{fig:ExpSetup_combined}(a). We use a red-detuned standing-wave and a blue-detuned running-wave nanofiber-guided light field to trap laser-cooled cesium atoms in two diametric linear arrays of potential minima in the evanescent field surrounding the nanofiber~\cite{LeKien04, Vetsch10}. The diameter of the nanofiber is $\SI{500}{\nano\meter}$, and the red-detuned  field has a free-space wavelength of $\SI{1064}{\nano\meter}$ and a total power of $\SI{1.96}{\milli\watt}$. The spatial period of the arrays is $\Delta z=\SI{498}{\nano\meter}$~\cite{LeKien04b}. The blue-detuned field has a power of \SI{17.8}{\milli\watt} and a free-space wavelength of \SI{763}{\nano\meter}. This reduces the Raman scattering background by a factor of 7 compared to previous settings~\cite{Vetsch10, Stolen84}. The light that is scattered by the atoms is collected with a microscope objective placed inside of the vacuum chamber (NA=0.29, working distance \SI{3.65}{\centi\meter})~\cite{Alt02}. An additional lens outside of the vacuum chamber (focal length \SI{10}{\centi\meter}) images the atoms onto a camera (Andor iXon Ultra 897). The magnification of the imaging system is $\sim 3$, and the point spread function (PSF) has a measured $1/e$-radius of $\sim\SI{10}{\micro\meter}$. A band-pass filter (Semrock $\SI{852}{\nano\meter}$ MaxLine) and a long-pass filter (Semrock $\SI{808}{\nano\meter}$ EdgeBasic) are used to further reduce the background. 

The atoms are loaded from a magneto-optical trap into the nanofiber-based trap using an optical molasses stage~\cite{Vetsch10}. This probabilistically populates both diametric arrays with at most one atom per trapping site~\cite{Schlosser01}. We remove atoms from the array behind the nanofiber using side-selective degenerate Raman heating~\cite{Meng18}. From the remaining ensemble, we only keep atoms that fall into the field of view of the imaging system. This is achieved by performing DRC with an external laser beam of \SI{200}{\micro\meter} diameter for $\SI{300}{\milli\second}$, i.e., for a time much longer than the 50-ms trap lifetime without cooling. Atoms outside of this range are thus lost. We then take images with a larger excitation beam-diameter of about \SI{1300}{\micro\meter}, thereby evenly illuminating the atoms (see SM). This beam propagates along the quantization axis ($+y$), is $\sigma^-$-polarized, and has a detuning of ~$-3\Gamma$ with respect to the D2 cycling transition. Here, $\Gamma$ is the linewidth of the D2 transition. In addition to generating scattered light for imaging, these settings are chosen to perform continuous DRC of the trapped atoms, thereby counteracting recoil and other heating mechanisms~\cite{Huemmer19}.

In Fig.~\ref{fig:ExpSetup_combined}(b), we present examples out of a series of images of nanofiber-trapped atoms that is obtained in one experimental run. In the upper panel (at time $t=0.585$~s after the start of the imaging series), two regions of pixels with a large number of counts are apparent. Each bright region shows light that stems from a single atom. The middle panel shows an image taken $\SI{0.39}{\second}$ later. One atom is still present while the other one has been lost from the trap. Even later ($t=1.95$~s), both atoms are lost (bottom panel). The series of images from which Fig.~\ref{fig:ExpSetup_combined}(b) was extracted is shown in the SM. We note that the 150-ms exposure time is shorter than the typical lifetime of the atoms when cooling is applied ($\tau_{\rm DRC}\approx$~\SI{1}{\second}). In Fig.~\ref{fig:ExpSetup_combined}(c), we analyze the same series of images by summing the relevant photon counts. The yellow (blue) symbols show this signal as a function of time, corresponding to the atom on the left (right). About~$180$ counts per exposure are observed for each atom, dropping to significantly smaller values when the atom is lost. 

We present a detailed analysis of the performance of our imaging procedure in the SM. For the case of $\SI{150}{\milli\second}$ integration time, an atom that is in the trap during the full exposure time is detect with a probability of $\sim 97.7~\%$. False detection, i.e., erroneously finding one or more atoms although there are no atoms in the trap, occurs in $\sim 7$~\% of the cases. Atoms can also be lost from the trap while imaging. We infer that the probability of detecting a trapped atom and then losing it by the end of the image integration time is $\sim 8~\%$. Moreover, we present images with only \SI{100}{ms} integration time including an animated image series as well as images taken with nanofiber-guided instead of external excitation light in the SM. Our analysis and the examples for different implementations of the imaging show that our method is reliable and can be applied in a variety of ways. The comparably large experimentally determined PSF of our imaging system limits the number of atoms that can be spatially resolved to be on the order of ten for the given section (\SI{\sim 300}{\micro\meter}) along the nanofiber over which atoms are prepared. This could be improved, e.g., by refined alignment of the imaging system and a microscope objective with larger NA.

\begin{figure}
	\includegraphics[width=0.9\columnwidth]{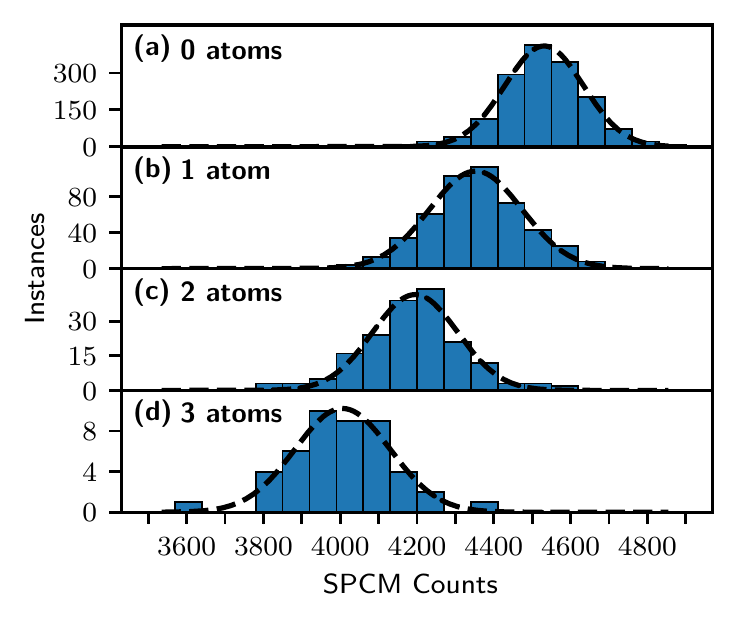}
	\caption{\textbf{Resonant transmission measurement with single-atom sensitivity}. (a)-(d) show histograms of the number of detected photons transmitted through the nanofiber for zero to three nanofiber-coupled atoms. Each atom absorbs $\sim 4$~\% of the guided light and, thus, the peak in the histogram shifts to the left with every additional atom. The integration time for each histogram is 90~ms, see main text. The count distributions are approximately Gaussian (see dashed lines for a fit).}
	\label{fig:reson_tm}
\end{figure}

We carry out two experiments that employ our capability of determining the number of atoms trapped along the nanofiber. First, we study the transmission of a resonant guided probe field as a function of the number of atoms in the trap. The transmitted light is detected with a single photon counting module (SPCM, Excelitas SPCM-AQRH-14-FC), see Fig.~\ref{fig:ExpSetup_combined}(a). In order to further reduce the background photon level arising from Raman scattering of the guided trapping laser fields, we filter the light that is sent to the SPCM using a combination of spectral filters as well as a narrow-band Fabry-P\'erot filter cavity. The combined transmission of the signal through the filtering system is $~50$~\%. The background SPCM count rate after filtering is $\approx$ \SI{2}{counts\,/\milli\second}. The nanofiber-guided probe light is quasi-linearly polarized in the plane of the trapped atoms and also performs DRC~\cite{Meng18}, thereby maximizing the interrogation time. However, it turns out that the DRC with resonant nanofiber-guided light is not quite as effective. Therefore, in order to still cool the atoms and to obtain good counting statistics for the resonant transmission measurement, we use an interleaved experimental scheme: We illuminate the atoms, alternating between the guided resonant probe light field and an external DRC laser beam, which is detuned by $-3\Gamma$. The probing lasts \SI{0.2}{\milli\second} and the cooling \SI{0.5}{\milli\second}, and we repeat the probing/cooling-cycle 450 times per experimental run. At the end of each experimental run, we also determine the rate of detected probe photons that are transmitted through the setup in the absence of atoms. This number is then used in order to correct for drifts. 

Figure~\ref{fig:reson_tm} shows the outcome of this transmission measurement. In (a), zero atoms are observed on the camera, and the histogram shows a distribution that is peaked at $\sim 4500$ SPCM counts. Panels~(b), (c), and (d) show the histograms for one, two, and three atoms observed on the camera, respectively. A shift of the SPCM count distribution to lower values by more than $100$ with every additional atom is apparent. We calculate the mean SPCM counts, $\bar{N}(i)$, for $i={0 \ldots 3}$ atoms. With each additional atom, we find that the mean extinction increases by $\{ \bar{\eta}(1),\bar{\eta}(2),\bar{\eta}(3) \}$={\{0.039(1), 0.039(1), 0.043(3)\}}, where $\bar{\eta}(i)=1-\bar{N}(i) /\bar{N}(i-1)$. The values are constant within the error and, thus, in agreement with Beer-Lambert law. The mean extinction per atom is consistent with the prediction for an atom--fiber surface distance of \SI{\sim 300}{\nano\meter}, which agrees well with the expected position of the trap minimum given the laser configuration used in our experiment.

\begin{figure}
	\includegraphics[width=1\columnwidth]{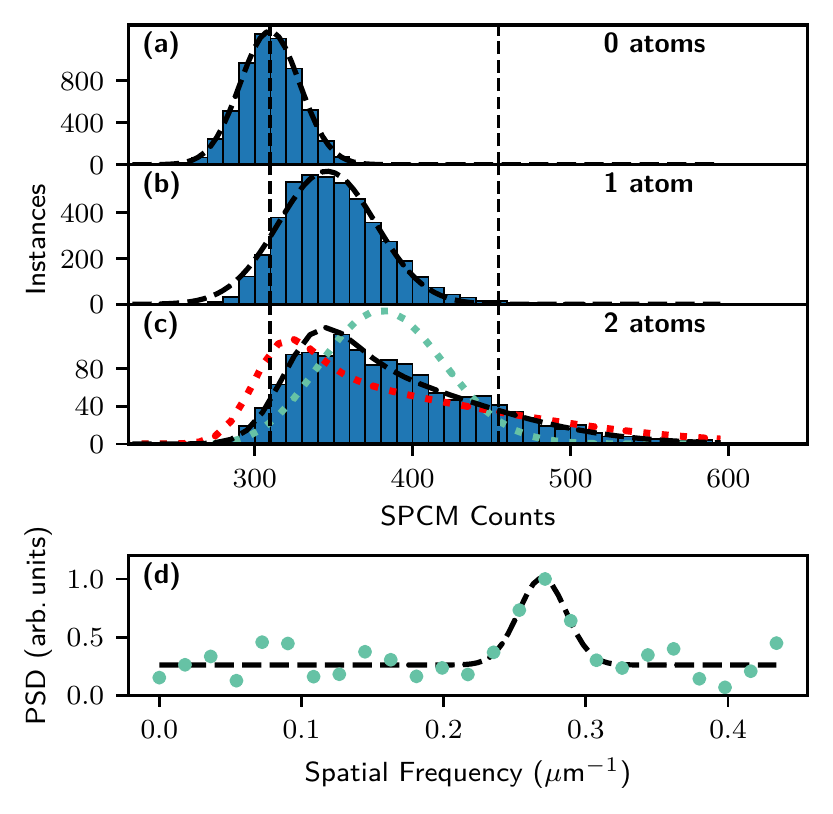}
	\caption{\textbf{Excitation of the nanofiber-guided mode by scattering light on nanofiber-coupled atoms}. The atoms are excited with a free-space laser beam, see Fig.~\ref{fig:ExpSetup_combined}(a). A fraction of the light scattered on the atoms enters the guided mode of the nanofiber. Histograms of the number of photons detected with the SPCM conditioned on having zero (a), one (b), and two (c) atoms detected in images taken with the camera. Panel~(a) shows a peak of the SPCM count distribution for zero atoms, indicating the background level. In (b), a peak that is shifted to larger values by about 35 counts compared to (a) is visible. This indicates that we can detect the nanofiber-coupled emission of single atoms. When two atoms emit into the nanofiber mode (in (c)), the respective light fields interfere with each other. As confirmed by the histogram data, the counts vary between the background level for perfectly destructive interference (vertical dashed line on the left) to four times the single-atom signal for constructive interference (line on the right). The black, red, and cyan lines show model predictions, also accounting for experimental imperfections. (d) Power spectral density (PSD) of the Fourier spectrum of the interference pattern deduced from the counts detected with the SPCM as a function of atom--atom distance determined using camera images. A clear peak at a spatial frequency of $\SI{0.269(3)}{\micro\meter^{-1}}$ is observed, consistent with our experimental setting. Parameters: Imaging duration 150~ms; detuning of excitation light from the D2 cycling transition $-3\Gamma$.
	}
	\label{fig:FC_coupled}
\end{figure}

In a second experiment, we study the scattering of an external light field by single as well as pairs of trapped atoms into the nanofiber-guided mode. Figure~\ref{fig:FC_coupled} shows histograms of the SPCM counts. In (a), no atom is observed on the camera. The corresponding peak in the histogram of the SPCM counts thus characterizes the background (a Gaussian fit yields $\bar{N}_{\rm bg}=309.63(9)$ counts, $\sigma_{N_{\rm bg}}=18.89(9)$ counts). Panel~(b) shows the histogram of the SPCM counts conditioned on the observation of one atom with the camera. Compared to (a), the peak is clearly shifted to higher values because of the additional light that the atom scatters into the nanofiber-guided mode. A Gaussian fit yields $\bar{N}_1=345.8(7)$ counts, $\sigma_{N_1}=30.9(7)$ counts. The width of the distribution in (a) is almost shot-noise limited, i.e., $\sqrt{N_{\rm bg}} \approx \sigma_{N_{\rm bg}}$. The width in (b), however, is about 70~\% larger than what is expected from Poissonian statistics. We attribute this to shot-to-shot variations of the atom--nanofiber coupling rate, of the intensity of the excitation laser beam (see SM), and of the transmission of the filter cavity because of frequency drifts. 

In Fig.~\ref{fig:FC_coupled}~(c), the light scattered into the nanofiber by two atoms is analyzed. We observe a photon count distribution which is qualitatively different from the one-atom case. For our parameters, we expect the scattering of the excitation light field by the atoms to be mainly coherent (saturation parameter $\sim 0.0023$), i.e., the scattered light fields will interfere constructively or destructively, depending on their relative phase. For a given fixed angle between the collimated excitation laser beam and the fiber axis, this relative phase is set by the atom--atom distance along the nanofiber. In the experiment, the relative phases of the scattered light fields are evenly sampled in the interval $\left[0, 2\pi\right)$, see SM. We now first assume that the amplitudes of the fields scattered by the two atoms onto the SPCM, though varying from shot to shot, are always equal to each other. Such common mode variations are to be expected, e.g., from drifts of the filter cavity, the excitation laser power, or the trapping potential. The red dotted line in Fig.~\ref{fig:FC_coupled}~(c) shows the predicted distribution of SPCM counts in this case. The distribution is normalized such that it has the same total number of instances as the experimental data. The underlying calculation takes into account the background signal inferred from (a) and the single-atom signal from (b). We attribute the deviation between the experimental data in (c) and this theory prediction to differential mode variations of the amplitudes of the fields scattered by the two atoms onto the SPCM. Such differential mode variations are to be expected from, e.g., an out-of-phase thermal motion of the two atoms in the radial direction of the trapping potential and a corresponding variation of the atom--nanofiber coupling strengths. In the extreme case of an on--off modulation of the coupling strengths, this leads to a distribution of SPCM counts, which effectively corresponds to the incoherent sum of the two single atom signals (green dotted line in (c)). Also this theory prediction deviates from the experimental data. Taking into account both common mode and differential mode variations and fitting their weight, we find very good agreement between the predicted distribution of SPCM counts and the experimental data for 71~\% common-mode and 29~\% differential-mode variations (black dashed line). 

In order to further analyze the measurement results that underlie the histogram in (c), we now determine the distance between the two atoms for each experimental run from three consecutive images. For this purpose, we vertically integrate images showing two atoms and then fit the result with the sum of two spatially offset one-dimensional Gaussians, yielding an estimate for the atom--atom distance. Assuming a Gaussian PSF with a 1/e-radius of 10~$\mu m$, see above, the statistical error of this distance estimation is about $\SI{0.8}{\micro\meter}$~\cite{alberti2016super}. We now examine the dependence of the SPCM counts on the atom--atom distance. To this end, we perform a Fourier transform of this signal and obtain the power spectrum shown in (d). A peak at a spatial frequency of $k_\text{exp}/2\pi=\SI{0.269(3)}{\micro\meter^{-1}}$, determined using a Gaussian fit, is visible. This corresponds to a predominant spatial modulation period of $\sim \SI{3.7}{\micro\meter}$, in good agreement with our expectations (see SM).

Summarizing, our demonstrated capability of in situ imaging and precision localization of single atoms significantly enriches the experimental toolbox for nanophotonic cold-atom systems. Here, beyond observing individual atoms in our nanofiber-based trap, it allowed us to perform an atom number-resolved measurement of the transmission of nanofiber-guided light and to thereby test Beer-Lambert's law atom by atom. Furthermore, our technique enabled us to study the collective scattering of light by two quantum emitters into a single-mode waveguide and to reveal the interference of the scattered fields. 

These results provide an excellent basis for future experimental studies of collective, waveguide-mediated effects~\cite{stannigel2012,Ramos14,Gonzalez-Tudela15,Shahmoon16}. For example, they may allow one to study sub- and superradiance~\cite{Solano17b} with an exactly known number of waveguide-coupled emitters. Moreover, imaging is an asset, for example, for the investigation of self-organization of quantum emitters along waveguides~\cite{Chang13,Griesser13}, including in the chiral domain~\cite{Holzmann14,Eldredge16}, and for the study of novel optical forces~\cite{Sukhov15,Rodriguez-Fortuno15,Scheel15,Kalhor16}. Single-site resolved imaging in nanophotonic cold atom systems such as the one studied here could become possible by using an imaging system with a larger NA, imaging light with a shorter wavelength, or by implementing super resolution techniques for trapped cold atoms~\cite{Subhankar19,McDonald19}. Finally, our results pave the way towards position-resolved real-time feedback and may enable the step-by-step assembly of cold atom-based nanophotonic quantum devices~\cite{Chang18,Samutpraphoot20}.

\begin{acknowledgments}
We thank D.~Weiss for contributions in the early stage of this project. Financial support from the Austrian Academy of Sciences (\"OAW, ESQ Discovery Grant Quantsurf), the
Austrian Science Fund (FWF, DK CoQuS, Project No.~W~1210-N16), and the European Union's Horizon 2020 research and innovation program under grant agreement No.~800942 (ErBeStA) is gratefully acknowledged.
\end{acknowledgments}
     
\bibliography{bibliography}
\clearpage

\end{document}



\title{Supplementary Material: Imaging and localizing individual atoms interfaced with a nanophotonic waveguide}

\author{Y.~Meng}
\affiliation{%
	Vienna Center for Quantum Science and Technology,\\
	TU Wien -- Atominstitut, Stadionallee 2, 1020 Vienna, Austria
}%
\affiliation{%
	Department of Physics, Humboldt-Universit\"at zu Berlin, 10099 Berlin, Germany
}%
\author{C.~Liedl}
\affiliation{%
	Department of Physics, Humboldt-Universit\"at zu Berlin, 10099 Berlin, Germany
}%
\author{S.~Pucher}

\affiliation{%
	Vienna Center for Quantum Science and Technology,\\
	TU Wien -- Atominstitut, Stadionallee 2, 1020 Vienna, Austria
}%
\affiliation{%
	Department of Physics, Humboldt-Universit\"at zu Berlin, 10099 Berlin, Germany
}%
\author{A.~Rauschenbeutel}
\affiliation{%
	Vienna Center for Quantum Science and Technology,\\
	TU Wien -- Atominstitut, Stadionallee 2, 1020 Vienna, Austria
}%
\affiliation{%
	Department of Physics, Humboldt-Universit\"at zu Berlin, 10099 Berlin, Germany
}%
\author{P.~Schneeweiss}
\affiliation{%
	Vienna Center for Quantum Science and Technology,\\
	TU Wien -- Atominstitut, Stadionallee 2, 1020 Vienna, Austria
}%
\affiliation{%
	Department of Physics, Humboldt-Universit\"at zu Berlin, 10099 Berlin, Germany
}%

\date{\today}
\maketitle

\onecolumngrid


\section{Atom detection}
\label{sec:atom_detection}
\subsection{Detection algorithm}
\label{subsec:imaging_loss}

\begin{figure}
	\includegraphics[width=0.5\columnwidth]{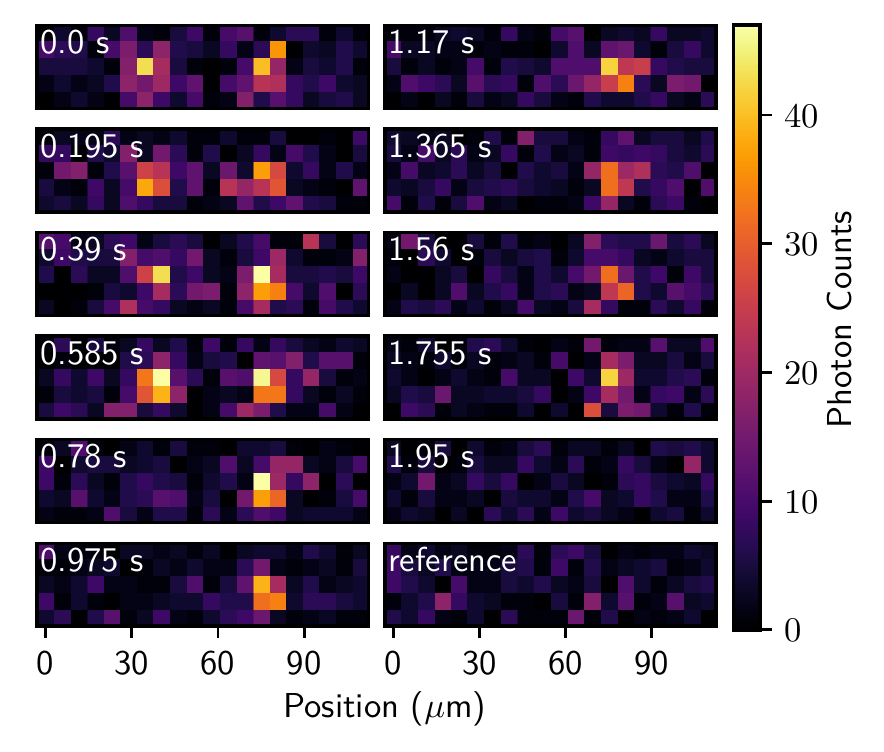}
	\caption{\textbf{Imaging of individual atoms with external excitation light.} An example of an image series, consisting of 11 images plus one reference image, is shown. The series is taken in one experimental run. The image integration time is \SI{150}{\milli\second}. There is a 45-ms waiting time between consecutive images, for technical reasons. The images are presented without background correction. The series of images shown here underlies Fig.~1(b) and (c) of the main manuscript.}
	\label{fig:series}
\end{figure}

We analyze images that were recorded with an integration time of 150~ms. The following procedure is executed in order to identify individual atoms in a given image: First, the image is convoluted with a discretized, two-dimensional Gaussian. The latter is set to zero outside an $11\times 11$ pixel region, and has a full width at half maximum (FWHM) of 1.5~pixels, or \SI{8.6}{\micro\meter} in the object plane. Next, a background image, generated by averaging many images without atoms, is subtracted. The camera is positioned such that the highest photon counts from a trapped atom, regardless of the position of its trapping site along the nanofiber, always falls into the same pixel row. Analyzing this row, we find the position and value of pixels that exhibit local maxima of the photon counts. If a pixel value of these local maxima in the convoluted, background-corrected image exceeds a threshold of 18~counts, we take this as a herald of a trapped atom and save the position of the pixel in order to analyze the same spot in consecutive images. For each experimental run, we take a series of 11 images plus one reference image with no atoms in the trap, see Fig.~\ref{fig:series} for an example series. A 45-ms waiting time between consecutive images is reserved to allow for the EMCCD camera to perform the image readout and to reinitialize for the next image. The analyzes outlined in Sects.~\ref{subsec:imaging_loss} and \ref{subsec:fidel_vs_thres} rely on one set of data that was obtained in 6000 experimental runs. At the beginning of each of these experimental runs, we load the nanofiber-based trap with $\sim 3$~atoms on average. Having a series of images for each experimental run allows us, for example, to process images conditioned on the outcome of the analysis of previous or following images. 

\begin{figure}
	\includegraphics[width=0.5\columnwidth]{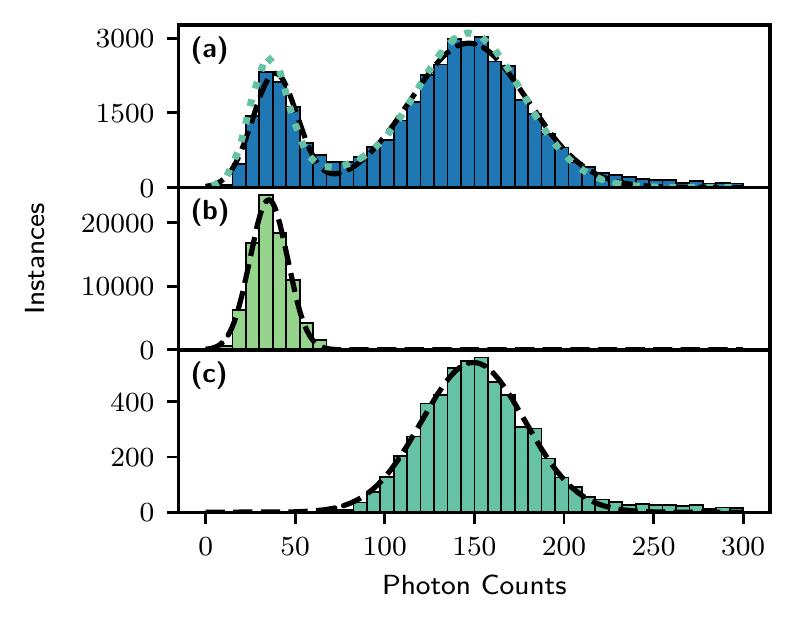}
	\caption{\textbf{Histogram of photon counts of single nanofiber-trapped atoms recorded with the camera.}  
	(a) Photon count histogram conditioned on a detection event in the previous image. The left and right peaks correspond to the background photon counts in the absence of atoms and the photon counts in the presence of one atom, respectively. (b) Photon count histogram with no atoms in the trap. A narrow background photon count distribution is visible. (c) Photon count histogram for single atoms when false detections and atom loss are excluded. The integration time is 150~ms for (a)-(c). For each series of 11 images, we use images number~$1, 3, 5, 7,$ and~$9$ for atom detection. See text for details.}
	\label{fig:histogram}
\end{figure}

In order to quantitatively characterize the performance of our detection procedure, we apply it to the data set detailed in the previous paragraph. We only take into account atom detection events in a $\sim \SI{300}{\micro\meter}$-long region of interest, see Sect.~\ref{sec:homogeneity}. This corresponds to the segment of the nanofiber where trapped atoms are prepared. If our procedure detects one and only one atom in the region of interest, we take the non-convoluted, non-background corrected version of the next image in the series and generate a histogram of the total photon counts in the $3\times 3$ pixel region that is centered on the pixel at which the atom was detected. This is repeated for all image series of the data set. The resulting histogram is shown in Fig.~\ref{fig:histogram}(a) and exhibits two distinct peaks, which we fit with the sum of two Gaussians, see black dashed line. The left peak is centered at 38.8(6) counts (standard deviation $\sigma=$~12.5(6) counts) while the right peak is centered at 146.9(8) counts ($\sigma=$~32.6(8) counts).

We expect that the left peak corresponds to the background photon counts in the absence of atoms. In order to check this assumption, we analyze the reference images, which were taken after briefly switching off the trap so that no atoms are present. We generate a photon count histogram of disjoint $3\times 3$ pixel regions in these images, see Fig.~\ref{fig:histogram}(b). Indeed, we observe a peak for similar values of photon counts as in Fig.~\ref{fig:histogram}(a). The black dashed line in Fig.~\ref{fig:histogram}(b) is a Gaussian fit, yielding a center value of 35.2(2) counts and a standard deviation of $\sigma=$10.5(2) counts. The presence of the left peak in Fig.~\ref{fig:histogram}(a) thus implies that, in some cases, no atom is present in the image following the one in which a detection event occurred. Two mechanisms can lead to such a false detection event. First, the detection algorithm may erroneously detect one or more atoms in an image when no atom is present. By running the detection algorithm on all 6000 reference images, we find that this indeed happens with $\sim 7$~\% probability. Second, due to the finite trap lifetime, an atom that is correctly detected in an image may be lost before the start of the integration time of the next image. 

The occurrence of both mechanisms can be ruled out by analyzing also the image directly following the one that used to generate the histogram. If an atom is found in this additional image, we obtain the photon count histogram in Fig.~\ref{fig:histogram}(c). There, only one peak is apparent which coincides with the right peak of the histogram in Fig.~\ref{fig:histogram}(a). This shows that this peak originates from the in-situ fluorescence of a single nanofiber-trapped atom. A Gaussian fit to the histogram in panel (c) (dashed black line) yields a center value of 149.4(5) counts ($\sigma=29.0(5)$ counts). Note that the threshold value of 18 differs significantly from the observed photon count values in Fig.~\ref{fig:histogram}(c): While the threshold condition refers to a single pixel, the histogram shows the total photon counts in $3\times 3$ pixel regions in the non-convoluted, non-background corrected images. Finally, the widths of both the right peak in panel (a) and the peak in panel (c) of Fig.~\ref{fig:histogram} are wider than what is expected for Poissonian counting statistics. This is, at least partially, explained by the dependence of the single-atom signal on the position at which the atom is observed on the camera, caused by vignetting and inhomogeneous illumination with the external excitation laser beam, see Sect.~\ref{sec:homogeneity}. 

Using the photon count distributions from panels (b) and (c), the independently measured trap lifetime during DRC of $\tau_{\rm DRC}\approx \SI{1}{\second}$, and the probability for false detections, we establish a model to predict the data shown in Fig.~\ref{fig:histogram}(a), see green dotted line. We obtain very good agreement, confirming that the histogram in panel (a) is indeed influenced by atom loss and false detection. From the model, we infer that the probability of not detecting a trapped atom due to its loss during the 150-ms image integration time is $\sim 6~\%$. Furthermore, we infer that the probability of detecting a trapped atom which is then lost by the end of the image integration time is $\sim 8~\%$. 
\newpage 
\subsection{Atom detection probability as a function of detection threshold}
\label{subsec:fidel_vs_thres}
\begin{figure*}[htb]
	\includegraphics[width=0.6\columnwidth]{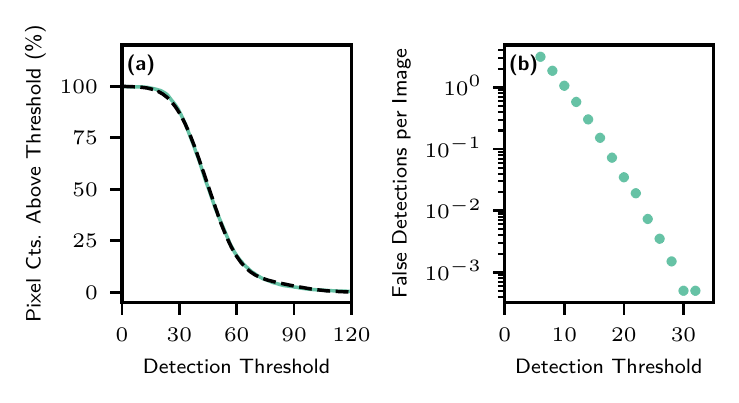}
	\caption{\textbf{Effect of threshold on atom detection.} (a)~
		Percentage of pixel counts above a given threshold value~(green solid line). We fit the sum of two cumulative Gaussian distribution functions to the data~(black dashed line). See text for details. (b) Mean number of false detections per image as a function of threshold. Here, we consider the 6000 reference images, i.e., images without atoms in the nanofiber-based trap. Note the different $x$-axes in panels (a) and (b): For panel (b), the detection procedure yields zero false detection events for a threshold value $\geq 34$. See text for details.}
	\label{fig:atom_detection_vs_threshold}
\end{figure*}

We examine the effect of the threshold on the atom detection. The basis for this analysis is the same data set that underlies Fig.~\ref{fig:histogram}, discussed in the previous section. Due to statistical fluctuations, background noise can result in pixel counts that exceed the threshold and, thus, lead to false atom detections. We can reduce the probability of false detections by increasing the detection threshold. However, a higher threshold will also decrease the atom detection efficiency. More quantitatively, one figure of merit is the probability of detecting a trapped atom for a given threshold. In order to infer this probability, we analyze the images that underlie Fig.~\ref{fig:histogram}(c). As discussed in Sect.~\ref{subsec:imaging_loss}, for these images, the detection of an atom in the previous and the next image with a fixed threshold of 18 ensures the presence of an atom in the trap during the entire integration time. As for the atom detection described in Sect.~\ref{subsec:imaging_loss}, we then convolute these images, subtract the background, and determine the pixel value at the position of the atom. The green solid line in Fig.~\ref{fig:atom_detection_vs_threshold}(a) shows the percentage of pixel counts above a threshold value as a function of this threshold. If all analyzed images contained exactly one atom, it would thus correspond to the efficiency of detecting a trapped atom. However, in some of the analyzed images, two nearby trapped atoms may mistakenly be detected as a single trapped atom and, in this case, an anomalously high pixel value will be found. This is apparent, e.g., in the small asymmetry of the histogram of photon counts shown in Fig. S2(c): The distribution closely resembles a Gaussian but falls off more slowly towards larger photon counts, a feature that we attribute to the occasional presence of two atoms. In order to infer the probability of detecting a single trapped atom in spite of this systematic error, we fit the sum of two cumulative Gaussian distribution functions to the data in Fig. S3(a). The two cumulative Gaussian distributions correspond to a single atom and to two unresolved trapped atoms, respectively, and we assume that the latter has on average twice the pixel counts of the former. The fit result is shown by the black dashed line in Fig. S3(a), which agrees very well with the data. From the fit, we estimate that the probability of detecting a single atom is $\sim 97.7 \%$ at a threshold value of 18. This threshold value is used for the data analysis of the results presented in the main text as well as for those in Sect.~\ref{subsec:imaging_loss}.

Another figure of merit is the number of false atom detections per image at a chosen threshold, see Fig.~\ref{fig:atom_detection_vs_threshold}(b). Here, we apply the atom detection procedure to the reference images, which were taken in the absence of trapped atoms. We calculate the mean number of false detections per image by averaging the occurrence of false atom detections over all 6000 reference images. From this data, we infer that the probability of at least one false detection to occur in a given image is $\sim 7$~\% at the chosen threshold of 18.

\newpage 
\section{Position dependence of single-atom signals}
\label{sec:homogeneity}
\begin{figure*}[htb]
	\includegraphics[width=0.5\columnwidth]{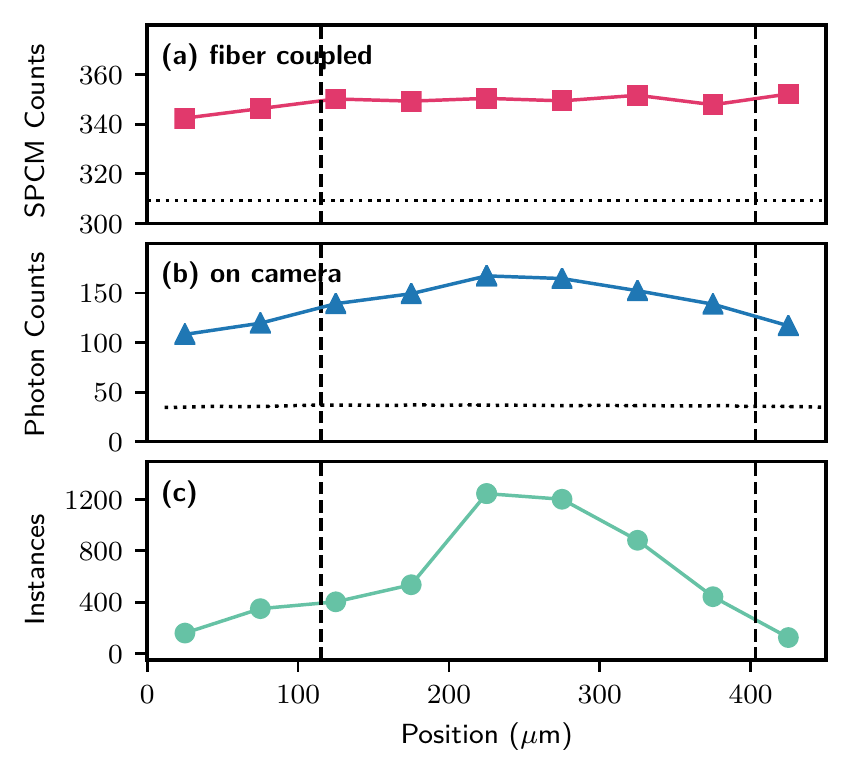}
	\caption{\textbf{Position dependence of single-atom signals.} (a) SPCM counts of nanofiber-coupled light. (b) Photon counts in a $3 \times 3$ pixel region detected with the camera. Panels~(a) and (b) show the average counts per atom per \SI{50}{\micro\meter}-interval. The horizontal dotted lines indicate the level of the background. (c) Distribution of the detected atom positions. We use bins of \SI{50}{\micro\meter} width and show instances of detected atom positions. The vertical dashed lines in panels~(a)-(c) indicate the boundaries of the region of interest. This region is used for the analysis underlying Fig.~3 of the main text as well as Figs.~\ref{fig:histogram} and \ref{fig:atom_detection_vs_threshold}. The lines connecting the symbols in panels~(a)-(c) are guides to the eye.}
	\label{fig:uniform_illum_external_light}
\end{figure*}

We analyze the level of the light scattered by a single nanofiber-trapped atom and recorded with the camera or detected with the SPCM as a function of the atom's position along the fiber. The analysis is performed on the data set described in Sect.~\ref{sec:atom_detection} and data that was recorded using the SPCM, in parallel to acquisition of each of the images in this data set. A possible variation of the excitation light intensity would affect both the camera and the SPCM signal. The latter would, in addition, depend on possible position-dependent variations of the coupling efficiency of the scattered light into the nanofiber. Other effects, like vignetting by the imaging optics, would solely influence the camera signal. In order to check a position dependence of the single-atom signals, we plot both the counts detected by the SPCM and the camera as a function of the atom position in panels~(a) and (b) of Fig.~\ref{fig:uniform_illum_external_light}, respectively. The photon counts in panel~(b) are summed over $3 \times 3$ pixel regions centered on the detected atom positions as before. The horizontal dotted line in panels~(a) and (b) indicates the background photon counts measured without atoms. Fig.~\ref{fig:uniform_illum_external_light}~(c) shows the occurrence of detected atom positions in the imaging region. We observe a bell-shaped distribution. We chose the position range between \SI{\sim 115}{\micro\meter} and \SI{\sim 403}{\micro\meter}, see vertical dashed lines, where most atoms are trapped, as our region of interest. In this region, the SPCM counts after background subtraction vary within only $\sim 5$~\%. The photon counts acquired with the camera are less uniform in the same region and vary by about $\sim 20$~\%. We mainly attribute this nonuniformity to imperfections of the imaging system. 

\newpage 
\section{Interference between nanofiber-coupled light fields scattered by two trapped atoms}
\begin{figure}[htb]
	\includegraphics[width=0.4\columnwidth]{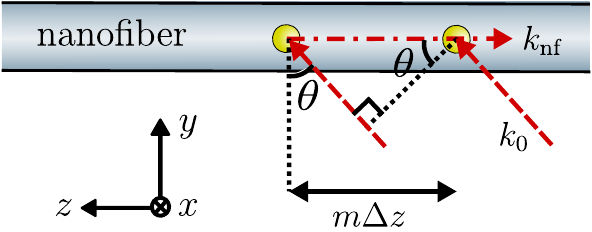}
	\caption{\textbf{Geometry involved in the scattering of external excitation light into the nanofiber.} The yellow circles represent two atoms. The lattice spacing of the nanofiber-based trapping potential is denoted $\Delta z$, and the atom-atom distance is an integer multiple, $m \Delta z$, of this lattice spacing. The incident angle of the excitation light field is denoted by $\theta$. The wavenumbers of the excitation light in free space and of the light guided in the nanofiber are denoted by $k_0$ and $k_{\rm nf}$, respectively. }
	\label{fig:supple_interference_drawing}
\end{figure}

\begin{figure}[htb]
	\includegraphics[width=0.5\columnwidth]{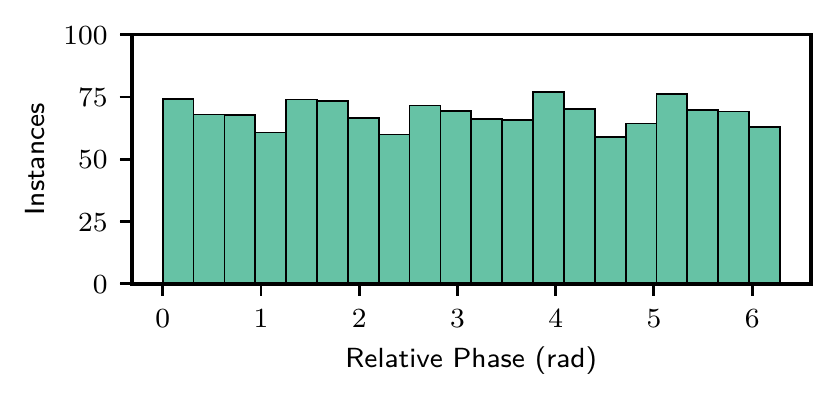}
	\caption{\textbf{Calculated histogram of relative phases between two nanofiber-coupled scattered light fields.}  }
	\label{fig:NF_couple_relative_phase}
\end{figure}

Fig.~\ref{fig:supple_interference_drawing} illustrates the setting and geometric path lengths relevant to the scattering of external excitation light into the nanofiber by two nanofiber-trapped atoms. The angle, $\theta$, between the wave vector of the external excitation light field and the $y$-axis of our coordinate system (see Fig.~1(a) of the main text) is about $16^ \circ$, i.e., far from the Bragg angle of about $35^\circ$ in our experiment. Given this incident angle and the long region of interest, we thus expect to evenly sample the interval $\left[0, 2\pi\right)$ with the relative phases between the two nanofiber-coupled light fields. Using the measured spatial distribution of the trapped atom positions, see Fig.~\ref{fig:uniform_illum_external_light}(c), we calculate the expected distribution of distances between two atoms. We then compute the relative phases between the scattered light fields from the inter-atomic distance using $\Delta\phi=m\Delta z (k_{\rm nf}+k_0\sin\theta$), where $m$ is the number of lattice sites between the two atoms and $\Delta z$ is the lattice spacing of the nanofiber-based trapping potential. From this, we generate the expected histogram of the relative phases, see Fig.~\ref{fig:NF_couple_relative_phase}. The flat distribution confirms our assumption of even sampling of the relative phase.

In order to obtain the spatial power spectrum shown in Fig.~3~(d) in the main text, we apply an additional constraint: When the distance between two trapped atoms is smaller than the point spread function diameter, they may erroneously be detected as one atom. To mitigate this effect, we only consider atom detection events with photon counts between 18 and 80. If the atom detection routine finds pixels with a photon count of more than 80 in the convoluted, background-corrected images, the detection event is discarded. This happens in $\sim 5$~\% of the detections and removes $\sim 73$~\% of the cases in which two atoms contribute to the detected peak while only eliminating $\sim 0.4$~\% of the single-atom events according to the model outlined in Sect.~\ref{subsec:fidel_vs_thres}. Furthermore, we only use data where the error of the fitted distance is smaller than 0.3 pixel, or \SI{1.7}{\micro\meter} in the object plane. In this way, we filter out images for which the atom positions are poorly determined with the Gaussian fit. Indeed, the point spread function can deviate from a Gaussian distribution, e.g., due to shot noise or imaging aberrations.

Based on these considerations, we analyze the results shown in Fig.~3(d) of the main manuscript, where we observe interference with a predominant spatial modulation period of $\sim \SI{3.7}{\micro\meter}=2\pi/k_\text{exp}$. This value is consistent with our experimental settings: For a plane wave incident under an angle $\theta$ with respect to the y-axis, see Fig.~\ref{fig:supple_interference_drawing} and Fig.~1(a) of the main manuscript, the condition for constructive interference of light scattered into the nanofiber is $d (k_0 \sin{\theta}+k_{\rm nf})=m\cdot 2\pi$ with $m \in \mathbf{N}$, where $d$ denotes the atom--atom distance, $k_{\rm 0}$ is the free-space wave number of the excitation light, and $k_{\rm nf}$ is the propagation constant of the nanofiber-guided light, respectively. If $d$ could be continuously varied, we would thus observe an interference signal with a spatial frequency of $1/\Delta d = (k_0 \sin{\theta}+k_{\rm nf})/2\pi$. In our experiment, the atom--atom distances can only take values which are integer multiples of the spatial period of the trapping potential, $\Delta z$. Since $1/(2\Delta z) < 1/\Delta d$, we thus undersample the interference signal and expect to observe a modulation at the spatial alias frequency, $k_{\text{alias}}/2\pi=|1/\Delta z-1/\Delta d|$. Inserting $k_{\text{alias}}=k_{\text{exp}}$ yields $\theta\approx 20^\circ$, in reasonable agreement with $\theta_\text{geom}\approx 16^\circ$, deduced from the geometry of our experimental setup.

\clearpage 
\newpage 
\section{Imaging of nanofiber-trapped atoms with 100~ms integration time}
\label{sec:fastimaging}
\begin{figure}[htb]
	\includegraphics[width=0.8\columnwidth]{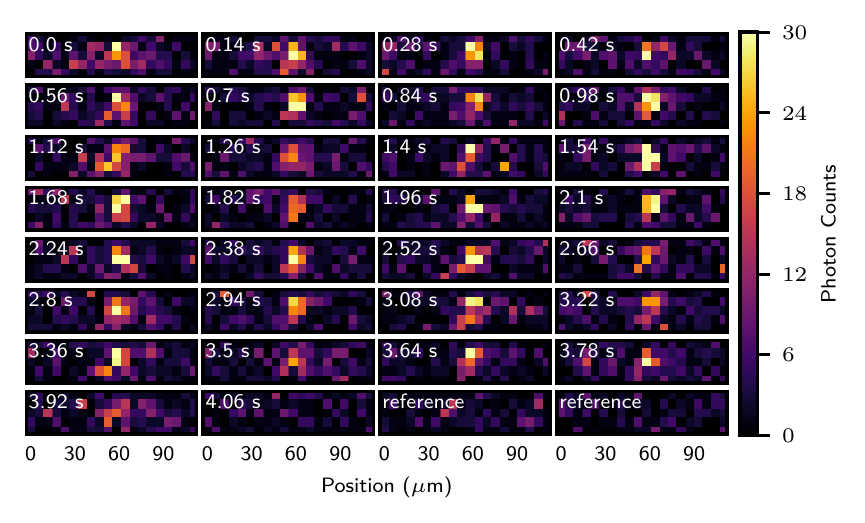}
	\caption{\textbf{Imaging of individual atoms with external excitation light.} The series consists of $30$ images separated by \SI{140}{\milli\second} as well as two reference images. No background correction was applied. The image integration time is \SI{100}{\milli\second}. The file \texttt{movie\_100ms\_exposure\_time.gif} of the supplementary material shows an animated series of those images.}
	\label{fig:real_time_imaging_100ms}
\end{figure}

Here, we image atoms with a shorter integration time of \SI{100}{ms}. For each experimental cycle, we record a series of $30$~images plus two reference images. An example series of raw atom images is shown in Fig.~\ref{fig:real_time_imaging_100ms}. There, we observe an atom for almost~\SI{4}{\second} until it is lost from the trap. A 40-ms waiting time between consecutive images is reserved to allow for the EMCCD camera to perform the image readout and to reinitialize for the next image. The external excitation beam is \SI{-3}{\Gamma} red-detuned with respect to the D2 cycling transition and $\sigma^-$-polarized.


\newpage 
\section{Atom imaging with nanofiber-guided excitation light}
\begin{figure*}[htb]
	\includegraphics[width=0.32\columnwidth]{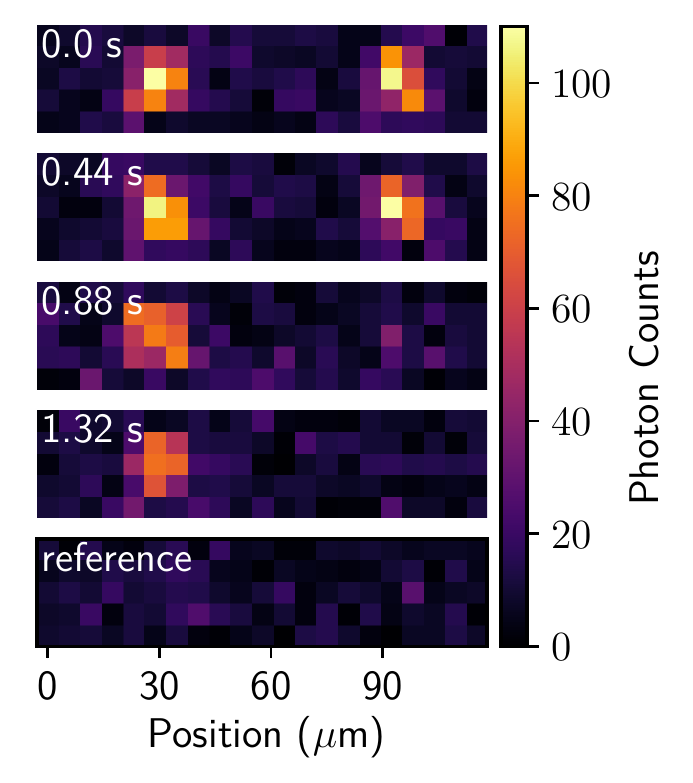}
	\caption{\textbf{Imaging of trapped atoms using nanofiber-guided excitation light.} The series consists of 4~images separated by \SI{440}{\milli\second} plus one reference image. No background correction was applied. The image integration time is \SI{400}{\milli\second}.}
	\label{fig:atom_images_FGD2}
	
\end{figure*}

Here, we demonstrate that our imaging technique can be implemented using nanofiber-guided instead of external excitation light. This approach is also applicable to other nanophotonic systems in which atom trapping is implemented with the evanescent field of the guided mode. The local polarization of the nanofiber-guided excitation light at the positions of the trapped atoms is almost perfectly $\sigma^-$-polarized~\cite{mitsch14}. As for atom imaging with external excitation light, the nanofiber-guided excitation light field implements degenerate Raman cooling, and we image the atoms by collecting the scattered photons during cooling. An example series of raw images with an integration time of \SI{400}{\milli\second} is presented in Fig.~\ref{fig:atom_images_FGD2}. There is a 40-ms waiting time between consecutive images, see Sect.~\ref{sec:fastimaging}. The nanofiber-guided excitation light is \SI{-3}{\Gamma} red-detuned with respect to the D2 cycling transition. 
\bibliography{bibliography}